\documentclass[journal]{IEEEtran}

\usepackage[utf8]{inputenc}
\usepackage{amsmath,amssymb,amsfonts}
\usepackage{algorithm}
\usepackage{algorithmic}
\usepackage{graphicx}
\usepackage{textcomp}
\usepackage{xcolor}
\usepackage{cite}
\usepackage{hyperref}
\usepackage{booktabs}

\usepackage{amsmath}
\usepackage{chngcntr}
\counterwithout{equation}{section}

\begin{document}

\title{A Hidden Markov Framework for Physically Interpretable Arc Stability Dynamics in Welding Systems}

\author{
\vspace{0.5em}
Hidir Selcuk Nogay 
\vspace{0.8em} \\

\small Bursa Uludag University, Faculty of Engineering, Department of Electrical and Electronics Engineering, Bursa, Türkiye
\vspace{0.5em} \\

\small \textit{Emails: hsnogay@uludag.edu.tr}
}


\maketitle

\begin{abstract}

Electric arc welding (EAW) exhibits strongly nonstationary and temporally evolving behavior, making reliable assessment of arc stability difficult using conventional frame-based approaches. In this study, arc dynamics are modeled as a sequence of latent operational regimes within a probabilistic state-space framework. The welding current signal is transformed into a time-frequency domain using Short-Time Fourier Transform (STFT), and a set of physically meaningful spectral descriptors, including energy, entropy, and centroid, is extracted to construct the observation sequence. A Hidden Markov Model (HMM) is employed to capture temporal dependencies and estimate the evolution of arc states. The analysis reveals three dominant regimes, transient, stable, and extinction, with a clear monotonic increase in spectral energy and a corresponding decrease in entropy, indicating reduced variability under stable conditions. Despite partial overlap in the feature space, the inferred state sequence exhibits strong temporal coherence, supported by high state persistence and low transition rates. These findings highlight the limitations of static classification and emphasize the importance of temporal modeling. The proposed framework provides an interpretable and physically consistent representation of arc behavior, enabling more realistic monitoring and analysis of stability dynamics in welding processes.

\end{abstract}


\section{Introduction}

\IEEEPARstart{E}{lectric} arc welding (EAW) is a complex industrial process governed by nonlinear interactions between electromagnetic fields, thermal dynamics, and plasma behavior \cite{tukahiruwa2023}. The stability of the electric arc directly determines weld quality, energy transfer efficiency, and structural integrity of the final product \cite{zhang2026}. However, the arc is inherently non-stationary and stochastic, exhibiting rapid temporal fluctuations due to plasma turbulence, impedance variations, and external disturbances. These characteristics make reliable monitoring and control of arc stability a challenging task, particularly in real-time industrial environments \cite{chen2026}.

Traditional approaches to arc monitoring rely on signal analysis techniques that extract descriptive features from the welding current \cite{akinci2010}. Time-frequency representations have shown strong capability in revealing localized spectral structures associated with underlying physical dynamics \cite{dealteriis2026}. Prior studies demonstrate that spectral characteristics derived from current signals can serve as informative indicators of system behavior across different domains, including power systems and biomechanical signal analysis \cite{akgun2018,kang2026, taskin2009}. In welding applications, these representations provide a physically meaningful description of energy redistribution within the arc, where variations in spectral energy reflect changes in plasma conductivity, arc length, and electromagnetic coupling \cite{khoshnaw2023, swierczynska2026}.

Building on this perspective, existing data-driven methods utilize time-frequency features with machine learning models to classify arc conditions into regimes such as transient, stable, and extinction states \cite{akinci2010, akinci2011, nogay2021}. However, these approaches rely on frame-level evaluation and treat observations independently, neglecting temporal dependencies. As demonstrated in Section IV, feature distributions exhibit partial overlap, indicating that static classifiers are inherently limited under realistic conditions. Consequently, such models provide instantaneous labeling rather than a continuous representation of arc behavior.

Another limitation concerns interpretability. Although descriptors such as energy and spectral entropy are widely used, their connection to physical arc dynamics is often not explicitly addressed. Similar challenges have been observed in related welding monitoring studies, where time-frequency and spectral features are employed for automated quality assessment, yet their direct linkage to underlying physical arc behavior remains limited \cite{garcia2026}. In practice, increasing energy combined with decreasing entropy reflects a more structured and constrained arc state, whereas higher entropy indicates increased stochastic fluctuations. Without embedding such physical interpretation, purely data-driven models remain limited in practical applicability \cite{akinci2010, taskin2009}.

Arc stability is inherently a temporally evolving process characterized by gradual transitions between different operating regimes, as also reflected in experimentally derived arc stability features under varying welding conditions \cite{kah2022}. This observation motivates the use of modeling frameworks that explicitly capture sequential dependencies over time. In this context, HMMs provide a natural probabilistic formulation, where the observed time--frequency features extracted via STFT are interpreted as emissions generated by an underlying sequence of latent states, a perspective consistent with recent process monitoring approaches that incorporate temporal dependencies and hidden state inference in industrial systems \cite{capezza2025, wang2026}.

In this study, arc dynamics are modeled as a stochastic process evolving over a finite set of hidden states. The welding current signal is transformed into a time-frequency domain using STFT, and a compact set of spectral descriptors is extracted to represent energy distribution and variability. These descriptors are treated as a sequential observation process within an HMM framework, enabling estimation of state transitions and reconstruction of arc evolution over time. Unlike conventional classification approaches, this formulation captures temporal persistence and transition dynamics, providing a more realistic representation of the welding process.

The main contributions of this work can be summarized as follows:

\begin{itemize}
\item A probabilistic state-space formulation of arc stability is developed, modeling the welding process as a sequence of hidden regimes.

\item A physically-informed observation model is constructed by linking STFT-based spectral features to underlying arc physics.

\item A temporal inference framework based on Hidden Markov Models is proposed to estimate state transitions and persistence behavior.

\item The study shifts arc analysis from static classification toward interpretable temporal modeling.
\end{itemize}

Accordingly, the contribution of this work lies in reformulating arc stability analysis as a temporal probabilistic inference problem grounded in both signal characteristics and physical interpretation.

The remainder of this paper is organized as follows. Section II presents the data acquisition and preprocessing steps. Section III introduces the Hidden Markov modeling framework. Section IV provides experimental evaluation and analysis. Section V concludes the study.

\section{Data Acquisition and Sequential Signal Representation}

The primary current signal acquired from an Electric Arc Welding Machine (EAWM) is modeled as a stochastic and non-stationary process in the Hilbert space $\mathcal{L}^2(\mathbb{R})$. The measured signal is expressed as:

\begin{equation}
i(t) = s(t) + n(t)
\label{eq:signal_model}
\end{equation}

As defined in (\ref{eq:signal_model}), the observed signal consists of the underlying arc dynamics $s(t)$ and an additive noise component $n(t)$. This formulation provides a physically meaningful interpretation in which the arc behavior is embedded within a stochastic observation model, explicitly accounting for measurement uncertainty and environmental disturbances.

To preserve temporal dependencies, the continuous-time signal is segmented into overlapping windows, forming a sequential representation:

\begin{equation}
\{i_t\}_{t=1}^{T}, \quad i_t \in \mathcal{L}^2(\mathbb{R})
\label{eq:sequence}
\end{equation}

As expressed in (\ref{eq:sequence}), each segment $i_t$ represents a localized observation of the arc process. This transformation enables the conversion of the raw signal into a structured temporal sequence suitable for probabilistic modeling.

A nonlinear feature mapping is then defined as:

\begin{equation}
\mathbf{v}_t = \Phi(i_t), \quad \mathbf{v}_t \in \mathbb{R}^d
\label{eq:feature_map}
\end{equation}

As indicated in (\ref{eq:feature_map}), the mapping $\Phi$ extracts informative descriptors from each segment, providing a compact representation that links physical signal characteristics with statistical inference.

\begin{figure*}[!t]
\centering
\includegraphics[width=\textwidth]{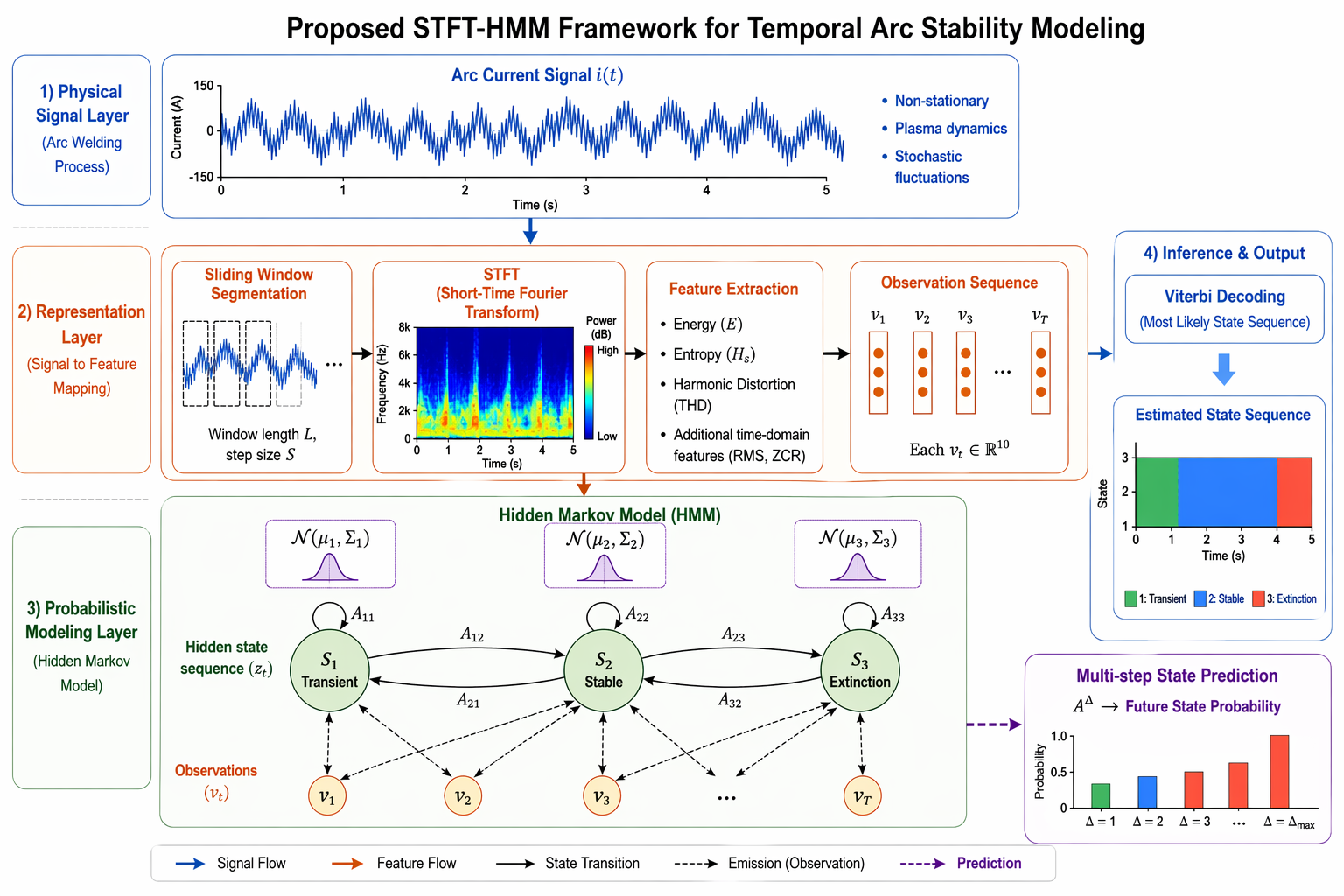}
\caption{Proposed STFT–HMM framework for temporal arc stability modeling, including signal acquisition, feature extraction, observation sequence construction, and probabilistic inference.}
\label{fig:architecture}
\end{figure*}

The overall acquisition and transformation process is illustrated in Fig.~\ref{fig:architecture}, which summarizes the complete signal-to-inference pipeline of the proposed framework. The diagram begins with the raw welding current signal $i(t)$, representing the physical arc behavior, which is inherently non-stationary and influenced by plasma dynamics and external disturbances.

In the first stage, the signal is segmented into overlapping time windows, enabling the preservation of temporal continuity while allowing localized analysis. Each segment is then transformed into the time–frequency domain using the STFT, producing a spectrogram that captures the instantaneous spectral energy distribution. This representation forms the basis for extracting physically interpretable features, including energy, spectral entropy, and centroid, which reflect both the intensity and structural organization of the arc.

These features are subsequently mapped into a low-dimensional observation space, forming a sequential feature vector $\mathbf{v}_t$. Since the resulting feature distributions exhibit partial overlap across different operating regimes, a Gaussian Mixture Model (GMM) is employed to provide an initial probabilistic partitioning of the observation space. This step enables the construction of a discrete observation sequence suitable for Hidden Markov modeling.

In the final stage, the HMM captures the temporal evolution of arc states through transition probabilities and emission distributions. The Viterbi algorithm is then applied to infer the most probable sequence of hidden states, providing a temporally consistent segmentation of the welding process into transient, stable, and extinction regimes.

Overall, Fig.~\ref{fig:architecture} highlights the integration of physical signal representation, statistical feature extraction, and probabilistic temporal modeling within a unified framework. This end-to-end structure not only enables interpretable analysis of arc dynamics but also establishes a direct connection between time–frequency characteristics and state-based inference.

\subsection{Experimental Setup and Measurement Infrastructure}

The experimental setup is based on a Metal Active Gas (MAG) welding platform operating under controlled conditions. The primary current is measured using a Hall-effect sensor (LEM series), which provides galvanic isolation, high linearity, and sufficient bandwidth for capturing transient arc behavior \cite{akinci2010}.

The measured signal can be expressed as:

\begin{equation}
i(t) = i_{\mathrm{arc}}(t) + n(t)
\label{eq:measurement}
\end{equation}

As described in (\ref{eq:measurement}), the observed signal is composed of the physical arc component $i_{\mathrm{arc}}(t)$ and measurement noise. This formulation ensures that the dominant structure reflects the underlying welding dynamics while maintaining robustness to noise.

The signal is digitized using a 16-bit data acquisition system with a sampling frequency satisfying:

\begin{equation}
f_s \gg 2 f_{\max}
\label{eq:nyquist}
\end{equation}

As expressed in (\ref{eq:nyquist}), this condition guarantees compliance with the Nyquist criterion, ensuring accurate capture of all relevant spectral components without aliasing. The selected sampling configuration provides sufficient temporal resolution to preserve both rapid transient variations and sustained arc behavior, forming a reliable basis for subsequent time--frequency analysis and temporal modeling.

\subsection{Sequential Characterization of Arc Regimes}

The welding process is modeled as a sequence of transitions between distinct operational regimes, rather than independent static classes. These regimes are defined as:

\begin{enumerate}
\item \textbf{Transient Phase:} Characterized by rapid ignition dynamics and broadband spectral content.
\item \textbf{Stable Phase:} Dominated by energy concentration around a fundamental frequency component.
\item \textbf{Extinction Phase:} Associated with irregular fluctuations and dispersed spectral energy.
\end{enumerate}

This formulation motivates the use of temporal modeling approaches, as the evolution of arc behavior is inherently sequential and cannot be accurately described by independent observations.

\subsection{Time--Frequency Transformation and Feature Extraction}

Each signal segment $i_t$ is analyzed using the Short-Time Fourier Transform:

\begin{equation}
S(t,f) = \int i(\tau) w(\tau - t) e^{-j2\pi f \tau} d\tau
\label{eq:stft}
\end{equation}

As defined in (\ref{eq:stft}), the STFT provides a localized time–frequency representation that captures transient variations in the spectral structure of the signal \cite{griffin1984, hammond1996}.

The corresponding energy distribution is given by:

\begin{equation}
P(t,f) = |S(t,f)|^2
\label{eq:energy}
\end{equation}

As expressed in (\ref{eq:energy}), $P(t,f)$ represents the instantaneous time–frequency energy density.

Based on this representation, physically interpretable features are defined.

The total spectral energy is computed as:

\begin{equation}
E(t) = \int P(t,f)\, df
\label{eq:total_energy}
\end{equation}

As defined in (\ref{eq:total_energy}), this quantity reflects the overall energy content of the arc and is directly related to power transfer characteristics.

Spectral entropy is used to quantify the complexity of the spectral distribution:

\begin{equation}
H_s(t) = - \int p(t,f) \log p(t,f)\, df
\label{eq:entropy}
\end{equation}

where $p(t,f)$ is the normalized spectral distribution. As indicated in (\ref{eq:entropy}), this feature captures the degree of disorder in the frequency domain and provides stability \cite{sucic2014, luo2012}.

The spectral centroid is defined as:

\begin{equation}
C(t) = \frac{\int f \cdot P(t,f)\, df}{\int P(t,f)\, df}
\label{eq:centroid}
\end{equation}

As expressed in (\ref{eq:centroid}), the centroid represents the dominant frequency location and its temporal evolution \cite{ghoraani2011}.

\subsection{Observation Space Construction for HMM}

The extracted features are combined to form a sequence of observation vectors:

\begin{equation}
\mathbf{V} = \{\mathbf{v}_1, \mathbf{v}_2, \dots, \mathbf{v}_T\}
\label{eq:obs_seq}
\end{equation}

As defined in (\ref{eq:obs_seq}), this sequence constitutes the input to the Hidden Markov Model.

Each observation vector is defined as:

\begin{equation}
\mathbf{v}_t = [E(t), H_s(t), C(t)]
\label{eq:feature_vector}
\end{equation}

As indicated in (\ref{eq:feature_vector}), the feature space is constructed using a compact set of physically interpretable descriptors. While these features capture key aspects of arc behavior, partial overlap between regimes is expected due to nonlinear dynamics and measurement variability.

This observation space provides a structured representation that enables the application of probabilistic temporal models, allowing the incorporation of both statistical properties and temporal dependencies in arc state estimation.

\section{Hidden Markov Modeling of Arc Stability Dynamics}

The temporal evolution of arc stability is formulated as a probabilistic sequence modeling problem, where the observable signal characteristics are governed by an underlying sequence of latent operational states. Unlike static classification approaches, this formulation explicitly captures temporal continuity, regime persistence, and transition dynamics.

Let the feature extraction operator defined in Section II be denoted as:

\begin{equation}
\Phi : \mathcal{L}^2(\mathbb{R}) \rightarrow \mathbb{R}^d
\label{eq:phi}
\end{equation}

As defined in (\ref{eq:phi}), the operator $\Phi$ maps the measured arc current signal into a structured feature space capturing energy, entropy, and spectral characteristics. This transformation enables the representation of complex plasma dynamics in a low-dimensional statistical domain.

The resulting observation sequence is expressed as:

\begin{equation}
\mathbf{v}_t = \Phi(i_t), \quad t = 1,2,\dots,T
\label{eq:obs}
\end{equation}

As shown in (\ref{eq:obs}), each time-localized signal segment is mapped into a feature vector, forming a sequential dataset suitable for temporal modeling. The relationship between the time–frequency representation and the resulting observation sequence is illustrated in Fig.~\ref{fig:architecture} .

\subsection{Latent State-Space Formulation}

The arc process is modeled as a discrete-time stochastic system with hidden states:

\begin{equation}
z_t \in \mathcal{S} = \{\text{Transient}, \text{Stable}, \text{Extinction}\}
\label{eq:states}
\end{equation}

Equation (\ref{eq:states}) defines the physically interpretable state space corresponding to different arc regimes. These states are not directly observable but inferred from the feature sequence.

The evolution of hidden states follows a first-order Markov process:

\begin{equation}
P(z_t \mid z_{t-1}) = A_{ij}
\label{eq:transition}
\end{equation}

As defined in (\ref{eq:transition}), the transition matrix $A$ encodes the probabilities of regime transitions. The Markov assumption simplifies temporal dependencies as:

\begin{equation}
P(z_t \mid z_{1:t-1}) = P(z_t \mid z_{t-1})
\label{eq:markov}
\end{equation}

Equation (\ref{eq:markov}) reflects the memoryless property, which is a reasonable approximation for short-time arc dynamics \cite{rabiner1989}.

\subsection{Observation Model and Emission Distribution}

The observation sequence $\{\mathbf{v}_t\}$ is assumed conditionally independent given the hidden states:

\begin{equation}
P(\mathbf{v}_t \mid z_t, \mathbf{v}_{1:t-1}) = P(\mathbf{v}_t \mid z_t)
\label{eq:cond_ind}
\end{equation}

As shown in (\ref{eq:cond_ind}), this assumption enables tractable inference by decoupling observations across time.

Each observation is modeled using a state-dependent Gaussian distribution:

\begin{equation}
P(\mathbf{v}_t \mid z_t = k) = \mathcal{N}(\boldsymbol{\mu}_k, \Sigma_k)
\label{eq:gaussian}
\end{equation}

Equation (\ref{eq:gaussian}) captures the statistical structure of each arc regime in the feature space. The separability of these distributions is empirically validated in Fig.~\ref{fig:architecture}, where partial overlap between states indicates the necessity of temporal modeling.

\subsection{Joint Probability Structure}

The joint probability of the hidden state sequence and observation sequence is defined as:

\begin{equation}
P(\mathbf{V}, Z) = \pi_{z_1} \prod_{t=2}^{T} A_{z_{t-1}, z_t} \prod_{t=1}^{T} P(\mathbf{v}_t \mid z_t)
\label{eq:joint}
\end{equation}

As expressed in (\ref{eq:joint}), the model integrates initial state probabilities, transition dynamics, and emission likelihoods into a unified probabilistic framework.

\subsection{Parameter Estimation}

The model parameters are defined as:

\begin{equation}
\lambda = \{A, \boldsymbol{\mu}_k, \Sigma_k, \pi\}
\label{eq:params}
\end{equation}

Equation (\ref{eq:params}) summarizes the complete parameter set.

These parameters are estimated by maximizing the likelihood:

\begin{equation}
\lambda^* = \arg\max_{\lambda} P(\mathbf{V} \mid \lambda)
\label{eq:mle}
\end{equation}

As defined in (\ref{eq:mle}), optimization is performed using the Baum–Welch algorithm \cite{rabiner1989}.

The posterior probabilities are:

\begin{equation}
\gamma_t(k) = P(z_t = k \mid \mathbf{V}, \lambda)
\label{eq:gamma}
\end{equation}

\begin{equation}
\xi_t(i,j) = P(z_t = i, z_{t+1} = j \mid \mathbf{V}, \lambda)
\label{eq:xi}
\end{equation}

Equations (\ref{eq:gamma}) and (\ref{eq:xi}) quantify state occupancy and transition expectations.

\subsection{State Inference and Decoding}

The most probable state sequence is obtained via:

\begin{equation}
\hat{Z} = \arg\max_{Z} P(Z \mid \mathbf{V}, \lambda)
\label{eq:viterbi}
\end{equation}

As defined in (\ref{eq:viterbi}), the Viterbi algorithm reconstructs the temporal evolution of arc regimes by identifying the most likely sequence of hidden states given the observed feature vectors \cite{rabiner1989}. The resulting state trajectory is analyzed in Section~IV, where both the temporal behavior and transition characteristics are presented.

The inferred sequence is expected to exhibit strong state continuity, reflecting the fact that arc dynamics evolve in a structured and temporally correlated manner rather than through random fluctuations. In particular, stable operating conditions persist over longer durations, while transient and extinction phases occur over shorter intervals associated with arc ignition and collapse.

From a modeling perspective, this behavior indicates that the Hidden Markov Model captures temporal dependencies and suppresses short-term variability in the feature space. The detailed transition characteristics and their statistical interpretation are provided in Section~IV.

Overall, the Viterbi-based decoding provides a coherent temporal interpretation of arc behavior, forming the basis for the experimental analysis presented in the subsequent section.

\subsection{Stability Metrics and Transition Analysis}

State persistence is defined as:
\begin{equation}
\text{Persistence}(k) = A_{kk}
\label{eq:persistence1}
\end{equation}

As defined in \eqref{eq:persistence1}, this metric quantifies the expected tendency of the system to remain in a given regime, directly reflecting temporal stability.

The instability transition probability is given by:
\begin{equation}
P_{\text{instability}} = P(z_{t+1} = \text{Extinction} \mid z_t = \text{Stable}
)
\label{eq:instability1}
\end{equation}

As expressed in \eqref{eq:instability1}, this quantity characterizes the likelihood of degradation from a stable operating condition toward arc extinction.

Multi-step prediction of future states is defined as:
\begin{equation}
P(z_{t+\Delta} \mid z_t) = A^{\Delta}
\label{eq:prediction1}
\end{equation}

As given in \eqref{eq:prediction1}, the transition matrix raised to the power $\Delta$ enables forecasting of arc behavior over multiple time steps, capturing long-term evolution patterns.

\subsection{Theoretical Properties}

\textit{Property 1 (Stationarity):}
\begin{equation}
\pi^* A = \pi^*
\label{eq:stationary}
\end{equation}

As defined in (\ref{eq:stationary}), the stationary distribution $\pi^*$ characterizes the long-term probabilistic behavior of the system, indicating the asymptotic proportion of time the arc process spends in each regime.

\textit{Property 2 (Monotonic Convergence):}
\begin{equation}
P(\mathbf{V} \mid \lambda^{(k+1)}) \geq P(\mathbf{V} \mid \lambda^{(k)})
\label{eq:monotonic}
\end{equation}

As expressed in (\ref{eq:monotonic}), the training procedure ensures non-decreasing likelihood during parameter estimation, providing a stable convergence behavior under the iterative optimization framework.

Additionally, clustering validity analysis in Fig.~\ref{fig:silhouette} confirms that, although feature distributions exhibit partial separability, temporal dependencies remain essential for robust state inference. This observation further justifies the use of the HMM framework over static clustering approaches, as it incorporates sequential structure into the modeling process.

\section{Experimental Results and Temporal Analysis}

The proposed framework is evaluated with respect to its ability to capture the temporal evolution of arc stability. The analysis integrates time–frequency representation, feature-space structure, and probabilistic temporal inference.

\subsection{Time–Frequency Observation Structure}

\begin{figure}[H]
\centering
\includegraphics[width=1.1\columnwidth]{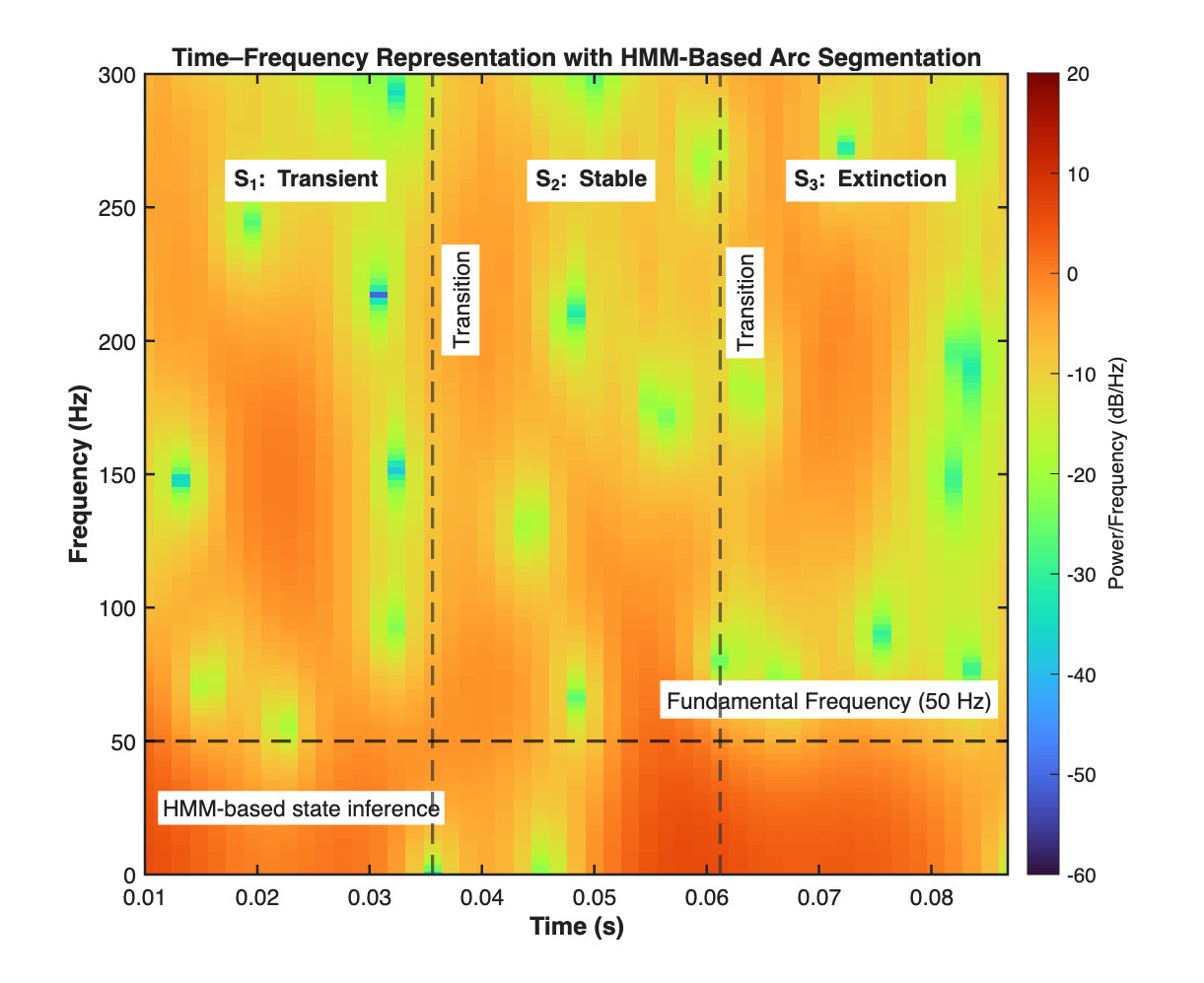}
\caption{Time–frequency representation of the welding current signal. Distinct spectral patterns corresponding to transient, stable, and extinction regimes are observable.}
\label{fig:stft}
\end{figure}

The STFT-based representation of the welding current signal is shown in Fig.~\ref{fig:stft}. This time–frequency map provides a localized spectral decomposition of the arc current, allowing the identification of dynamic energy redistribution patterns associated with different operational regimes. The transient region is characterized by highly dispersed spectral content, where energy spreads across a wide frequency band, reflecting unstable plasma ignition and irregular arc formation. In contrast, the stable regime exhibits a more concentrated and continuous energy distribution, particularly around the fundamental frequency and its harmonics, indicating sustained arc conduction and relatively uniform plasma dynamics.

As the process transitions toward extinction, the spectral structure becomes less coherent, with intermittent high-intensity bursts and localized frequency components, suggesting arc constriction and breakdown of stable energy transfer. The horizontal band around the fundamental frequency (approximately 50–60 Hz) remains a persistent feature; however, its relative dominance varies across regimes, providing a key indicator of arc stability. From a statistical perspective, this representation reveals non-stationary and time-dependent feature distributions, where both spectral variance and energy concentration evolve across time segments.

Importantly, the vertical transition boundaries highlighted in Fig.~\ref{fig:stft} indicate regime changes that are not sharply separable in the instantaneous feature space but emerge clearly in the time–frequency domain. These observations directly motivate the construction of observation vectors used in the Hidden Markov Model, where each time frame is represented by features derived from the local spectral energy distribution. Thus, the STFT representation serves not only as a visualization tool but as the foundational layer for feature extraction, enabling the HMM to capture temporal dependencies and state transitions inherent in the welding process. This establishes a direct link between the physical signal behavior, its statistical representation, and the probabilistic modeling framework proposed in this study.

The localized energy operator is defined as:
\begin{equation}
E_{\mathcal{B}}(t) = \int_{\mathcal{B}} |S(t,f)|^2 df
\label{eq:energy_local}
\end{equation}

As expressed in (\ref{eq:energy_local}), the spectral energy distribution evolves continuously over time, indicating that arc dynamics are inherently sequential.

\subsection{Temporal Evolution of Spectral Features}

\begin{figure}[H]
\centering
\includegraphics[width=1.1\columnwidth]{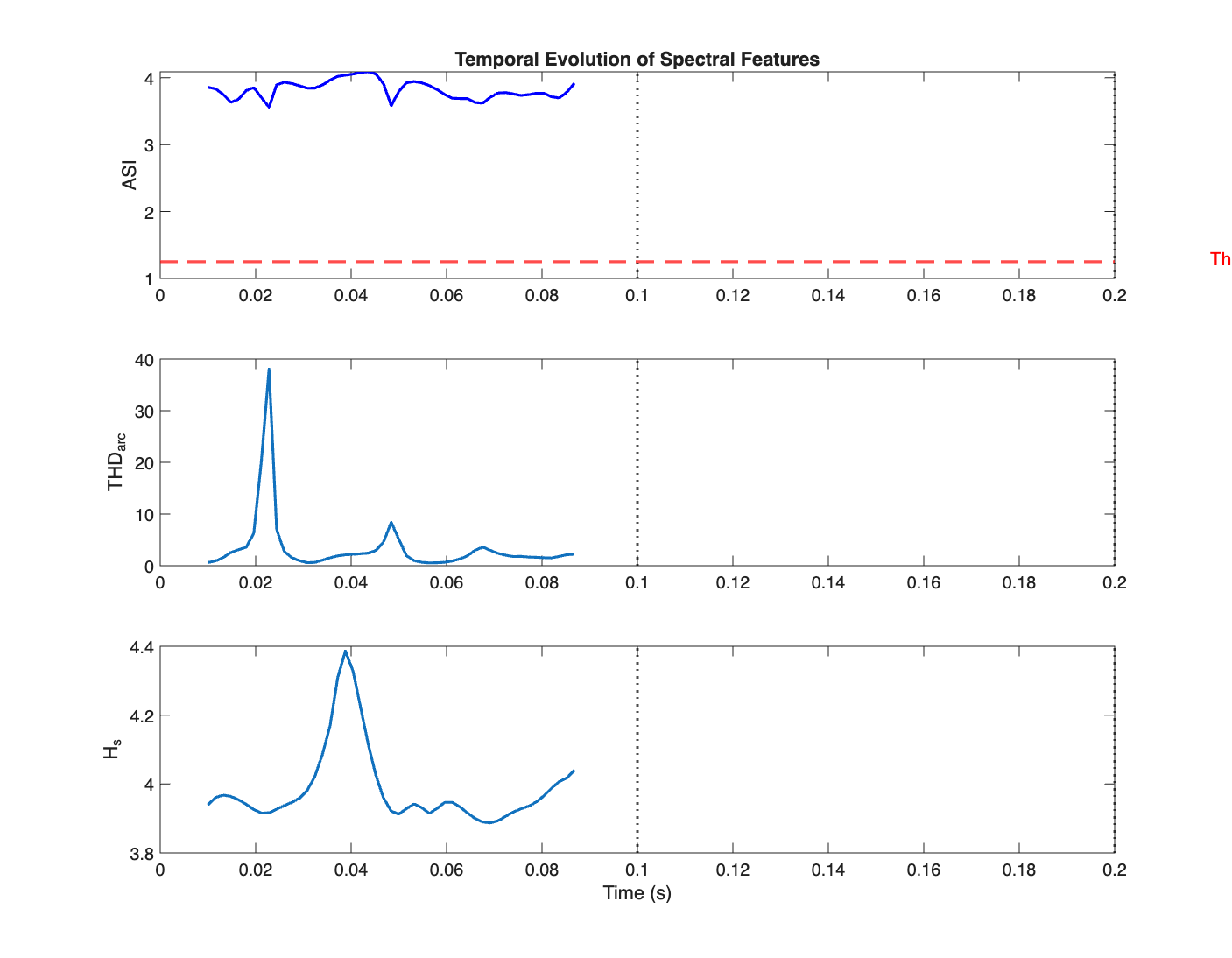}
\caption{Temporal evolution of spectral features including energy-based index, harmonic distortion, and entropy.}
\label{fig:features_time}
\end{figure}

The temporal evolution of key spectral descriptors is illustrated in Fig.~\ref{fig:features_time}. The figure presents the time-varying behavior of energy-related and distribution-based features, revealing non-stationary patterns that cannot be adequately captured by static analysis. In particular, the entropy profile exhibits a pronounced peak around $t \approx 0.04$~s, indicating a temporary increase in spectral dispersion. This behavior corresponds to a highly irregular arc condition, where energy is distributed across a broader frequency range, reflecting increased plasma instability.

Simultaneously, the distortion-related component shows sharp transient spikes, suggesting the presence of localized nonlinear phenomena and harmonic amplification during this interval. These abrupt changes are not sustained, but instead occur intermittently, highlighting the stochastic nature of arc dynamics. In contrast, the remaining segments of the signal demonstrate relatively smoother variations, indicating quasi-stationary behavior associated with more stable arc conditions.

From a statistical perspective, the temporal trajectories of these features reveal heteroscedastic behavior, where both the variance and distribution characteristics change over time. This violates the implicit assumptions of stationary feature distributions commonly used in conventional classification frameworks. Moreover, the presence of short-duration peaks embedded within otherwise stable regions implies that instantaneous feature values are insufficient for reliable state discrimination.

The vertical transition marker in Fig.~\ref{fig:features_time} further emphasizes that feature distributions shift across time intervals, suggesting the existence of underlying regime changes. These transitions are not abrupt in the feature space but occur gradually, reinforcing the need for a sequential modeling approach. In this context, the observed temporal correlations and structured variability directly motivate the use of Hidden Markov Models, where state persistence and transition probabilities can capture the evolution of arc behavior.

Fig.~\ref{fig:features_time} provides critical evidence that arc dynamics are governed by temporally evolving spectral characteristics rather than static patterns. This observation forms the basis for integrating time–frequency feature extraction with probabilistic state inference, enabling a more realistic and physically consistent modeling of welding processes.

These trajectories show gradual variations rather than abrupt changes, reinforcing the necessity of temporal modeling.

\subsection{Feature Space Structure and Separability}

\begin{figure}[H]
\centering
\includegraphics[width=1.1\columnwidth]{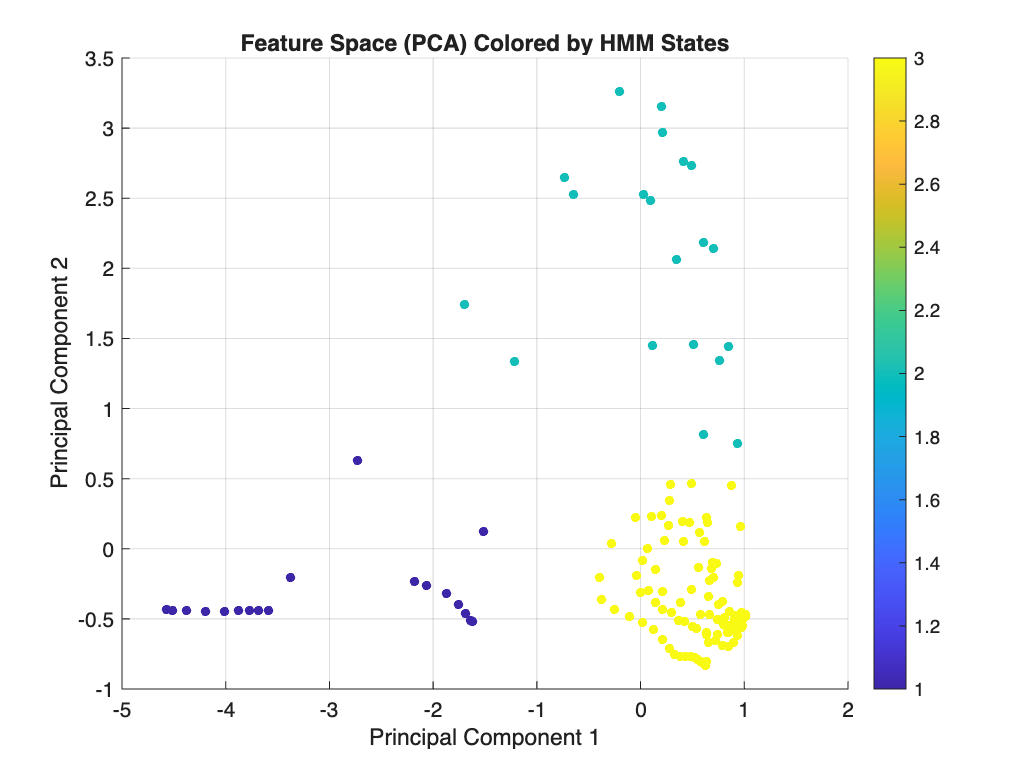}
\caption{PCA projection of feature vectors colored by inferred HMM states. Partial overlap between clusters is evident.}
\label{fig:pca}
\end{figure}

The PCA-projected feature space is shown in Fig.~\ref{fig:pca}. The projection reveals a structured yet partially overlapping distribution of the extracted feature vectors, reflecting the intrinsic complexity of arc welding dynamics. Three dominant clusters can be visually identified, corresponding to the operational regimes inferred by the HMM framework.

The transient regime is located in the negative region of the first principal component and exhibits a wide spread, indicating significant variability in feature values. This behavior reflects unstable ignition dynamics, where energy is low and spectral distribution remains dispersed. In contrast, the stable regime forms a more compact cluster with moderate energy and entropy levels, indicating reduced variability and quasi-stationary arc behavior. The extinction regime appears as a relatively well-separated cluster, characterized by higher energy concentration and lower entropy, suggesting a more structured yet constrained spectral pattern associated with arc termination.

Despite this overall structure, partial overlap between clusters is observed, particularly between transient and stable states. This indicates that instantaneous feature vectors alone are insufficient for reliable state discrimination, as transitional behaviors produce ambiguous observations in the feature space. Such overlap highlights the limitation of static clustering approaches and motivates the use of temporal modeling frameworks.

From a statistical perspective, the first principal component primarily captures energy-related variations, while the second component reflects entropy and spectral distribution characteristics. This alignment between statistical structure and physical interpretation supports the relevance of the selected feature set.

\begin{figure}[H]
\centering
\includegraphics[width=1.1\columnwidth]{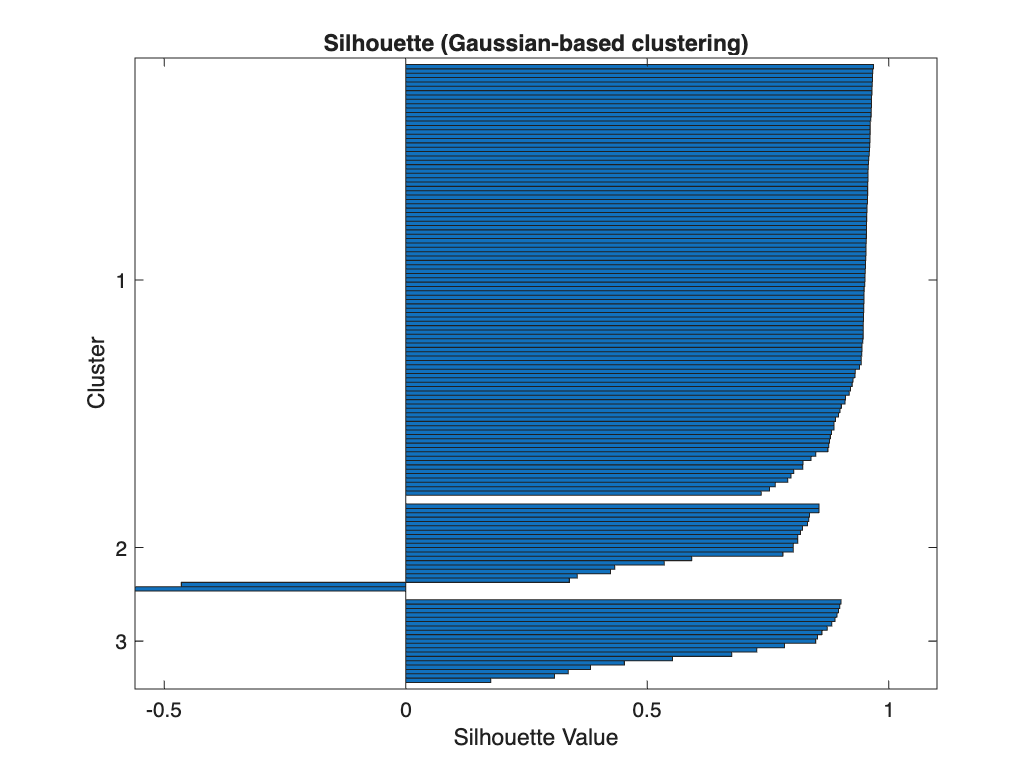}
\caption{Silhouette analysis of clustering structure. Overlap regions indicate limitations of static clustering.}
\label{fig:silhouette}
\end{figure}

To further quantify the degree of cluster separability and validate these observations, silhouette analysis is introduced in Fig.~\ref{fig:silhouette}.

Cluster separability is further evaluated using silhouette analysis, as shown in Fig.~\ref{fig:silhouette}. The silhouette coefficient measures how well each observation fits within its assigned cluster relative to neighboring clusters, providing a quantitative assessment of clustering quality. Values close to one indicate well-separated samples, whereas values near zero or negative reflect overlap or potential misclassification. As observed in Fig.~\ref{fig:silhouette}, a significant portion of the samples exhibit high silhouette values, confirming the presence of distinguishable structure in the feature space. However, the existence of samples with low or negative values indicates partial overlap between clusters, particularly near regime boundaries. This behavior is consistent with the underlying physical nature of arc welding, where transitions between operating regimes occur gradually rather than abruptly. Therefore, the silhouette analysis not only validates the overall clustering tendency but also highlights the limitations of purely static partitioning, reinforcing the necessity of incorporating temporal modeling to achieve robust state inference.

Cluster separability is further quantified using silhouette analysis in Fig.~\ref{fig:silhouette} \cite{rousseeuw1987}.

The class-conditional statistics are defined as:

\begin{equation}
\boldsymbol{\mu}_k = \mathbb{E}[\mathbf{v}_t | z_t = k], \quad 
\Sigma_k = \mathrm{Cov}(\mathbf{v}_t | z_t = k)
\label{eq:stats}
\end{equation}

As defined in \eqref{eq:stats}, the mean vector $\boldsymbol{\mu}_k$ and covariance matrix $\Sigma_k$ characterize the statistical distribution of feature vectors associated with each hidden state. These parameters capture both the central tendency and variability of the observation space, enabling a compact representation of state-specific feature behavior.

The emission model is given by:

\begin{equation}
P(\mathbf{v}_t | z_t = k) = \mathcal{N}(\boldsymbol{\mu}_k, \Sigma_k)
\label{eq:emission}
\end{equation}

As expressed in \eqref{eq:emission}, the observation vectors are modeled using a Gaussian distribution conditioned on the hidden state. This formulation assumes that each regime generates features with a distinct statistical structure, allowing probabilistic differentiation between states based on their underlying distributions.

As observed in Fig.~\ref{fig:pca} and Fig.~\ref{fig:silhouette}, these distributions are not fully separable, demonstrating that static classification is insufficient and highlighting the need for temporal modeling approaches \cite{rabiner1989}.

\subsection{State-Based Feature Statistics}

The statistical properties of each state are summarized in Table~\ref{tab:state_stats}.

\begin{table}[H]
\centering
\caption{State-based Feature Statistics}
\label{tab:state_stats}
\begin{tabular}{lcccccc}
\toprule
State & $E_{\text{mean}}$ & $E_{\text{std}}$ & $H_{\text{mean}}$ & $H_{\text{std}}$ & $C_{\text{mean}}$ & $C_{\text{std}}$ \\
\midrule
Transient  & 664.68 & 955.60 & 2.58 & 0.74 & 2434.47 & 83.76 \\
Stable     & 1365.88 & 916.25 & 2.01 & 0.15 & 2500.75 & 6.93 \\
Extinction & 2752.16 & 745.06 & 1.84 & 0.05 & 2504.10 & 6.39 \\
\bottomrule
\end{tabular}
\end{table}

Table~\ref{tab:state_stats} provides a quantitative summary of the feature distributions associated with each inferred arc regime. The reported values are presented in their original scales, preserving their physical meaning while enabling direct interpretation of energy, entropy, and frequency-related characteristics.

A clear progression is observed across the states. The transient regime exhibits lower mean energy and significantly higher variability, reflecting unstable ignition conditions and irregular arc behavior. In contrast, the stable regime shows more consistent feature values with reduced standard deviations, indicating a relatively controlled and quasi-stationary process. The extinction regime is characterized by the highest energy concentration and minimal variability in entropy, suggesting a more deterministic yet physically constrained state preceding arc collapse.

From a statistical perspective, the reduction in entropy variance across states indicates a gradual decrease in spectral complexity, while the increase in energy reflects a stronger concentration of localized spectral components. Additionally, the relatively small variance in centroid values for the stable and extinction regimes implies that dominant frequency components remain confined within a narrow band, reinforcing the structured evolution of the arc.

Overall, these results demonstrate that although feature distributions exhibit partial overlap in the global feature space, their statistical properties differ systematically across regimes. This structured variation supports the validity of the proposed state-based representation and highlights the importance of combining statistical descriptors with temporal modeling for accurate characterization of arc dynamics.

\subsection{State Transition Dynamics}

The transition probability matrix is given in Table~\ref{tab:hmm_transition}, which provides a compact probabilistic representation of how the welding process evolves between different operational regimes over time. Unlike static clustering outputs, this matrix explicitly encodes temporal dependencies, revealing not only dominant states but also the likelihood and direction of regime transitions.

\begin{table}[H]
\centering
\caption{Estimated HMM Transition Matrix}
\label{tab:hmm_transition}
\begin{tabular}{c|ccc}
\toprule
 & Transient & Stable & Extinction \\
\midrule
Transient  & 0.95 & 0.00 & 0.05 \\
Stable     & 0.00 & 0.90 & 0.10 \\
Extinction & 0.04 & 0.10 & 0.87 \\
\bottomrule
\end{tabular}
\end{table}

As shown in Table~\ref{tab:hmm_transition}, the diagonal elements dominate the matrix, indicating that once the system enters a given regime, it tends to remain in that state for a sustained period. This persistence behavior is formally defined by the self-transition probability in \eqref{eq:persistence}, where each diagonal term represents the likelihood of remaining in the same state between consecutive time steps.

The diagonal elements correspond to:
\begin{equation}
A_{kk} = P(z_{t+1}=k \mid z_t=k)
\label{eq:persistence}
\end{equation}

As expressed in \eqref{eq:persistence}, high values of $A_{kk}$ directly reflect temporal stability within each regime. In particular, the transient and stable states exhibit strong persistence (0.95 and 0.90), while the extinction state also remains highly stable (0.87), indicating that arc behavior evolves in a continuous and structured manner rather than through random fluctuations. From a physical standpoint, this behavior is consistent with the inertia of plasma dynamics, where established arc conditions tend to persist over short time intervals.

The instability transition probability is defined as:
\begin{equation}
P_{\text{inst}} = P(z_{t+1} = \text{Extinction} \mid z_t = \text{Stable})
\label{eq:instability}
\end{equation}

As defined in \eqref{eq:instability}, this metric captures the probability of degradation from a stable operating condition toward arc extinction. According to Table~\ref{tab:hmm_transition}, this transition occurs with a probability of approximately 0.10, indicating a measurable but relatively limited risk of instability under normal operating conditions. This provides a quantitative interpretation of arc failure likelihood and establishes a direct connection between probabilistic modeling and physical system reliability.

Together, \eqref{eq:persistence} and \eqref{eq:instability} characterize both regime continuity and critical transition pathways. While \eqref{eq:persistence} reflects how long the system remains stable, \eqref{eq:instability} identifies the key transition associated with performance degradation. This dual interpretation highlights that the proposed HMM framework not only segments the welding process but also enables a deeper understanding of stability dynamics, which cannot be captured by frame-based analysis alone.

\subsection{Temporal State Inference}

\begin{figure}[H]
\centering
\includegraphics[width=1.1\columnwidth]{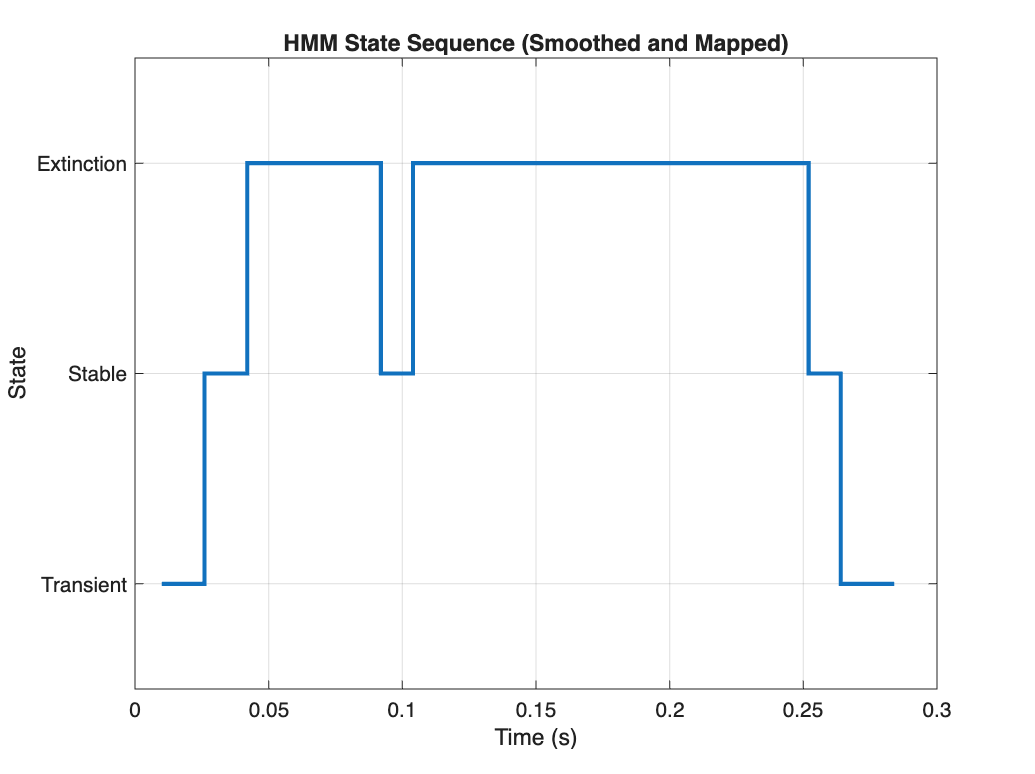}
\caption{Temporal evolution of hidden states obtained via Viterbi decoding.}
\label{fig:hmm}
\end{figure}

The inferred state sequence is shown in Fig.~\ref{fig:hmm}.

The model captures smooth transitions between regimes, consistent with physical arc behavior.

\subsection{Prediction of Instability Evolution}

Future state probabilities are computed as:
\begin{equation}
P(z_{t+\Delta} | z_t) = A^{\Delta}
\label{eq:prediction}
\end{equation}

Instability detection is defined as:
\begin{equation}
P(z_{t+\Delta} = \text{Extinction} | z_t) > \theta
\label{eq:threshold}
\end{equation}

\begin{table}[H]
\centering
\caption{Temporal Validation Metrics of the Proposed HMM Framework}
\label{tab:table3}
\begin{tabular}{lc}
\toprule
Metric & Value \\
\midrule
Average State Persistence & 0.8802 \\
Transition Entropy & 1.2625 \\
Mean State Duration & 19.7143 \\
State Duration Variance & 616.9048 \\
Transition Ratio & 0.0435 \\
\bottomrule
\end{tabular}
\end{table}

The temporal behavior of the inferred arc states is quantitatively evaluated using the metrics summarized in Table~\ref{tab:table3}. These metrics are designed to assess the stability, consistency, and dynamic evolution of the state sequence generated by the proposed Hidden Markov Model.

As reported in Table~\ref{tab:table3}, the high average state persistence (0.8802) indicates that the model effectively captures stable arc regimes with strong temporal continuity, avoiding rapid and unrealistic switching between states. This observation is further supported by the low transition ratio (0.0435), which demonstrates that state transitions occur infrequently and in a controlled manner, consistent with physical arc behavior.

The mean state duration of approximately 20 frames confirms that each detected regime corresponds to a meaningful temporal segment rather than noise-driven fluctuations. Although the state duration variance is relatively high, this result reflects the inherent nature of the welding process, where stable operation persists over extended intervals while transient and extinction phases occur over shorter durations.

Furthermore, the transition entropy value (1.2625) suggests that while the system exhibits structured transitions, it retains a degree of stochastic variability, which is expected in real-world arc dynamics. Overall, Table~\ref{tab:table3} provides quantitative evidence that the proposed framework successfully models the temporal structure of arc stability, going beyond static feature-based analysis and enabling a physically consistent interpretation of regime evolution.

\subsection{Discussion}

The obtained results consistently indicate that arc stability should be interpreted as a continuous temporal process rather than a set of independent observations. The gradual transitions observed between regimes and the strong state persistence values confirm that arc behavior evolves in a structured and temporally correlated manner.

The overlap observed in the feature space (Fig.~\ref{fig:pca}) provides direct evidence that conventional static classification approaches are inherently limited, particularly in nonstationary environments where temporal dependencies play a critical role \cite{pavlovic1999, haykin1998}. Similar observations have been reported in time-varying signal analysis studies, where feature distributions fail to form clearly separable clusters due to transitional dynamics \cite{holan2018, kohlmorgen2004}. This limitation becomes particularly critical in arc welding processes, where regime boundaries are not sharply defined but evolve progressively under changing physical conditions.

The temporal validation metrics presented in Table~\ref{tab:table3} further reinforce this interpretation. The high average state persistence and low transition ratio indicate that the inferred states exhibit strong temporal coherence, while the non-negligible transition entropy reflects the stochastic nature of real-world arc dynamics. These findings are consistent with prior studies emphasizing the importance of temporal modeling in nonstationary signal environments \cite{sin1995, basseville1983}.

From a physical perspective, the observed behavior aligns with the underlying plasma dynamics of arc welding. Stable arc conditions persist over extended intervals due to sustained energy transfer, whereas transient and extinction phases occur over shorter and more irregular durations. This explains both the high persistence values and the variability in state durations, highlighting the limitations of rigid, frame-based interpretations.

By integrating spectral descriptors derived from time--frequency representations with probabilistic temporal modeling, the proposed framework captures both the local frequency characteristics and their temporal evolution. This dual representation enhances interpretability by linking signal features to physically meaningful regimes, while also enabling predictive insight through structured state transitions.

Overall, the results demonstrate that incorporating temporal dependencies is not merely beneficial but necessary for realistic modeling of arc behavior. The proposed approach therefore provides a more faithful representation of regime evolution compared to static feature-based methods, particularly in systems characterized by gradual transitions and nonstationary dynamics.

\section{Conclusion}

This study investigates electric arc welding dynamics through a temporal modeling perspective by combining spectral descriptors derived from time--frequency representations with a Hidden Markov Model. Instead of treating observations as independent snapshots, the proposed approach models arc behavior as an evolving process, allowing the underlying structure of regime transitions to be explicitly captured.

The results show that arc operation can be consistently described using three regimes: transient, stable, and extinction. The extracted features exhibit a clear trend, where spectral energy increases while entropy decreases across states. This pattern suggests that the system becomes more constrained as it approaches instability, reflecting the structured yet nonlinear nature of arc behavior.

Analysis in the feature space indicates that these regimes are not sharply separable. The overlap observed in PCA projections and clustering analysis confirms that static, frame-based methods are insufficient to describe the process. In contrast, the temporal model produces coherent state sequences, supported by strong persistence in the transition structure and smooth evolution over time.

The estimated transition dynamics further indicate that state changes occur gradually rather than abruptly, aligning with the physical behavior of the welding process. This provides a more faithful interpretation of arc stability compared to purely feature-based classification schemes.

The framework therefore offers an interpretable link between signal characteristics and dynamic behavior, with potential use in monitoring and early detection of instability. Future work will focus on improving labeling strategies, expanding the dataset under different operating conditions, and evaluating alternative temporal models to strengthen generalization and predictive capability.


\begin{thebibliography}{1}


\bibitem{tukahiruwa2023}
G. Tukahiruwa and C. Wandera,
"Influence of process parameters in gas-metal arc welding (GMAW) of carbon steels,"
in \emph{Welding of Metallic Materials},
IntechOpen, 2023, doi: 10.5772/intechopen.1002730.


\bibitem{zhang2026}
L. Zhang, L. Chen, B. Jia, S. Yang, M. Pan, and H. Pan,
"Interpretable arc stability monitoring via Physics-Guided Data Programming in robotic gas metal arc welding,"
\emph{Journal of Manufacturing Processes}, vol. 166, pp. 352--369, May 2026, doi: 10.1016/j.jmapro.2026.03.049.

\bibitem{chen2026}
L. Chen, M. Liu, X. Liu, Z. Yu, P. Duan, Y. Liu, M. Fa, and B. Zheng,
"Bidirectional LSTM-based compensation of potential-asymmetry induced errors in double probe plasma diagnostics,"
\emph{Journal of Physics D: Applied Physics}, vol. 59, no. 3, p. 035206, Jan. 2026, doi: 10.1088/1361-6463/ae357e.


\bibitem{akinci2010}
T. C. Akinci, "Time-frequency analysis of the current measurement by hall effect sensors for electric arc welding machine," \emph{Mechanika}, vol. 85, no. 5, pp. 66--71, 2010.

\bibitem{dealteriis2026}
G. de Alteriis, R. Schiano Lo Moriello, A. Astarita, and A. T. Silvestri,
"A coherence-driven method for instability frequency identification and uncertainty evaluation in friction stir welding,"
\emph{Measurement}, vol. 275, p. 121385, May 2026, doi: 10.1016/j.measurement.2026.121385.


\bibitem{kang2026}
S. Kang, T. Yang, S. Jeon, M. Kang, and J. Shin,
"Accurate real-time weld shape monitoring in laser welding using spectral features and a deep neural network,"
\emph{Measurement}, vol. 270, p. 120903, Apr. 2026, doi: 10.1016/j.measurement.2026.120903.

\bibitem{akgun2018}
O. Akgun, A. Akan, H. Demir, and T. C. Akinci, "Analysis of Gait Dynamics of ALS Disease and Classification of Artificial Neural Networks," \emph{Tehnički vjesnik}, vol. 25, no. S1, pp. 183--187, 2018. doi: 10.17559/TV-20171011153544.


\bibitem{taskin2009}
S. Taskin, S. Seker, M. Karahan, and T. C. Akinci, "Spectral analysis for current and temperature measurements in power cables," \emph{Electric Power Components and Systems}, vol. 37, no. 4, pp. 415--426, 2009. doi: 10.1080/15325000802599383.

\bibitem{khoshnaw2023}
F. Khoshnaw, I. Krivtsun, and V. Korzhyk,
"Arc welding methods,"
in \emph{Welding of Metallic Materials: Methods, Metallurgy, and Performance},
Elsevier, 2023, pp. 37--71, doi: 10.1016/B978-0-323-90552-7.00004-3.


\bibitem{swierczynska2026}
A. Swierczynska, A. Janeczek, C. Pandey, et al.,
"A bibliometric review of A-TIG welding: unveiling global research trends,"
\emph{International Journal of Advanced Manufacturing Technology}, vol. 143, pp. 1367--1387, Mar. 2026, doi: 10.1007/s00170-026-17474-2.

\bibitem{akinci2011}
T. C. Akinci, H. S. Nogay, and G. Gokmen, "Determination of optimum operation cases in electric arc welding machine using neural network," \emph{Journal of Mechanical Science and Technology}, vol. 25, no. 4, pp. 1003--1010, 2011. doi: 10.1007/s12206-011-0202-9.

\bibitem {nogay2021}
H. S. Nogay and T. C. Akinci, "Classification of operation cases in electric arc welding machine by using deep convolutional neural networks," \emph{Neural Computing and Applications}, vol. 33, pp. 6657--6670, 2021. doi: 10.1007/s00521-020-05443-4.

\bibitem{garcia2026}
A. García Rodríguez, C. C. Barriga Castellanos, J. E. Rocha-Gonzalez, and E. Bárcenas,
"Time--frequency and spectral analysis of welding arc sound for automated SMAW quality classification,"
\emph{Sensors}, vol. 26, no. 8, p. 2357, 2026, doi: 10.3390/s26082357.


\bibitem{kah2022}
P. Kah, G. O. Edigbe, B. Ndiwe, et al.,
"Assessment of arc stability features for selected gas metal arc welding conditions,"
\emph{SN Applied Sciences}, vol. 4, p. 268, 2022, doi: 10.1007/s42452-022-05150-5.

\bibitem{capezza2025}
C. Capezza, A. Lepore, and K. Paynabar,
"Stream-based active learning for process monitoring,"
\emph{Technometrics}, vol. 68, no. 1, pp. 159--171, 2026, doi: 10.1080/00401706.2025.2561744.

\bibitem{wang2026}
Z. Wang, S. He, X. Zhao, and M. Zhen,
"Condition-based maintenance for production systems: Inferring degradation states from quality characteristics,"
\emph{IISE Transactions}, 2026, doi: 10.1080/24725854.2026.2628678.

\bibitem{griffin1984}
D. Griffin and J. Lim,
"Signal estimation from modified short-time Fourier transform,"
\emph{IEEE Transactions on Acoustics, Speech, and Signal Processing},
vol. 32, no. 2, pp. 236--243, Apr. 1984, doi: 10.1109/TASSP.1984.1164317.

\bibitem{hammond1996}
J. K. Hammond and P. R. White,
"The analysis of non-stationary signals using time-frequency methods,"
\emph{Journal of Sound and Vibration},
vol. 190, no. 3, pp. 419--447, 1996, doi: 10.1006/jsvi.1996.0072.

\bibitem{sucic2014}
V. Sucic, N. Saulig, and B. Boashash,
"Analysis of local time-frequency entropy features for nonstationary signal components time supports detection,"
\emph{Digital Signal Processing}, vol. 34, pp. 56--66, Nov. 2014, doi: 10.1016/j.dsp.2014.07.013.


\bibitem{luo2012}
G. Luo, D. Zhang, Y. Koh, K. Ng, and W. Leong,
"Time--frequency entropy-based partial-discharge extraction for nonintrusive measurement,"
\emph{IEEE Transactions on Power Delivery}, vol. 27, no. 4, pp. 1919--1927, Oct. 2012, doi: 10.1109/TPWRD.2012.2200911.


\bibitem{ghoraani2011}
B. Ghoraani and S. Krishnan,
"Time--frequency matrix feature extraction and classification of environmental audio signals,"
\emph{IEEE Transactions on Audio, Speech, and Language Processing}, vol. 19, no. 7, pp. 2197--2209, Sep. 2011, doi: 10.1109/TASL.2011.2118753.

\bibitem{rabiner1989}
L. R. Rabiner, "A tutorial on hidden Markov models and selected applications in speech recognition," \emph{Proceedings of the IEEE},
vol. 77, no. 2, pp. 257--286, Feb. 1989, doi: 10.1109/5.18626.

\bibitem{rousseeuw1987}
P. J. Rousseeuw, "Silhouettes: A graphical aid to the interpretation and validation of cluster analysis," \emph{Journal of Computational and Applied Mathematics},
vol. 20, pp. 53--65, Nov. 1987, doi: 10.1016/0377-0427(87)90125-7.


\bibitem{pavlovic1999}
V. Pavlovic, B. J. Frey, and T. S. Huang,
"Time-series classification using mixed-state dynamic Bayesian networks,"
in \emph{Proc. IEEE Conf. Computer Vision and Pattern Recognition (CVPR)},
Fort Collins, CO, USA, 1999, pp. 609--615, doi: 10.1109/CVPR.1999.784983.


\bibitem{haykin1998}
S. Haykin and D. J. Thomson,
"Signal detection in a nonstationary environment reformulated as an adaptive pattern classification problem,"
\emph{Proceedings of the IEEE},
vol. 86, no. 11, pp. 2325--2344, Nov. 1998, doi: 10.1109/5.726792.

\bibitem{holan2018}
S. H. Holan and N. Ravishanker,
"Time series clustering and classification via frequency domain methods,"
\emph{WIREs Computational Statistics},
vol. 10, no. 6, p. e1444, 2018, doi: 10.1002/wics.1444.


\bibitem{kohlmorgen2004}
J. Kohlmorgen,
"Tracking and visualization of changes in high-dimensional non-parametric distributions,"
in \emph{Proc. IEEE Workshop on Machine Learning for Signal Processing (MLSP)},
Sao Luis, Brazil, 2004, pp. 203--212, doi: 10.1109/MLSP.2004.1422975.


\bibitem{sin1995}
B. Sin and J. H. Kim,
"Nonstationary hidden Markov model,"
\emph{Signal Processing},
vol. 46, no. 1, pp. 31--46, Sep. 1995, doi: 10.1016/0165-1684(95)00070-T.


\bibitem{basseville1983}
M. Basseville and A. Benveniste,
"Sequential segmentation of nonstationary digital signals using spectral analysis,"
\emph{Information Sciences},
vol. 29, no. 1, pp. 57--73, Feb. 1983, doi: 10.1016/0020-0255(83)90009-9.



\end{thebibliography}
\end{document}